\documentclass{raa_twocolumn}
\usepackage{graphicx,times, dblfloatfix}             
\usepackage{natbib}
\usepackage{amssymb,amsmath}
\bibpunct{(}{)}{;}{a}{}{,}
\usepackage[pagebackref=true]{hyperref}

\begin{document}

  \title{The coded mask of the ECLAIRs telescope onboard the SVOM space mission}

   \volnopage{Vol.0 (202x) No.0, 000--000}      
   \setcounter{page}{1}          

   \author{C. Lachaud 
      \inst{1,*}\footnotetext{$*$Corresponding Author.}
   \and A. Givaudan
      \inst{1}
   \and  M. Karakac \inst{1} \and W. Bertoli \inst{1} \and S. Dheilly \inst{1} \and C. Juffroy \inst{1} \and C. Chapron \inst{1} \and A. Gros \inst{2,**} \and S. Schanne \inst{2} \and S. Begoc \inst{3} \and P. Guillemot
      \inst{3} \and H. Pasquier \inst{3} 
   }

   \institute{Université Paris Cité, CNRS, Astroparticule et Cosmologie, F-75013 Paris, France {\it cyril.lachaud@in2p3.fr}\\
        \and
             CEA Paris-Saclay, Institut de Recherche sur les lois Fondamentales de l'Univers, 91191 Gif sur Yvette, France; ** retired;\\
        \and
             CNES, 18 Avenue Edouard Belin, 31401 Toulouse cedex 9, France\\
\vs\no
   {\small Received 202x month day; accepted 202x month day}}

\abstract{ ECLAIRs is a hard X-ray coded-mask telescope onboard the SVOM space mission, designed to detect and localize high-energy transients, in particular gamma-ray bursts. Operating over the 4–150~keV energy range, ECLAIRs extends coded-mask imaging to an unusually low-energy threshold. Achieving sensitivity down to 4~keV while maintaining performance up to 150~keV motivated the development of a novel self-supporting coded mask. This design addresses both scientific and mechanical challenges through dedicated pattern-generation algorithms and an innovative stiffened sandwich structure. We present the rationale, development, and final implementation of the ECLAIRs coded mask.
\keywords{mission: SVOM ---  instruments: ECLAIRs, coded mask, X-ray}
}

   \authorrunning{C. Lachaud et al}            
   \titlerunning{The coded mask of ECLAIRs onboard SVOM}  
   \maketitle
   
\section{Introduction}           
\label{sect:intro}

The \textit{ECLAIRs} instrument~(\cite{godet25}) onboard the \textit{SVOM} space mission~(\cite{cordier25}) is a hard X-ray telescope operating in the 4--150~keV energy range.
It employs the coded-mask imaging technique~\citep[see][for a detailed overview]{golgro22}, which is particularly well suited to this domain when a very large field of view (2 sr, as in the case of ECLAIRs) is required, since such coverage cannot be achieved with grazing-incidence mirrors or lobster-eye optics.

The relatively low energy threshold of 4~keV is especially challenging for such a system, as conventional supporting structures cannot be used—since they would absorb low-energy photons passing through the mask apertures—and this constraint has motivated the development of a novel design, which we describe in this article.

For clarity, only a simplified overview of the development is presented here, although in practice the mechanical and pattern design aspects were strongly interdependent and have changed during the development phase.

\section{Coded mask constraints for an innovative telescope}
\label{sect:Constraints}

ECLAIRs is a hard X-ray telescope designed to survey a large fraction of the sky for the detection and localization of transient high-energy phenomena, such as Gamma-Ray Bursts (GRBs), Tidal Disruption Events (TDEs), and Active Galactic Nuclei (AGN) outbursts. Its field of view (FoV) spans $89 \times 89\,\deg ^2$, corresponding to approximately one sixth of the sky, with localization accuracy of a source at detection limit required to be better than $11.5\,\mathrm{arcmin}$ at the 90\% confidence level (including a margin of $1.5\,\mathrm{arcmin}$). The localization accuracy requirement is driven by the SVOM narrow field of view instruments, especially the VT which gives the highest constraint, since its FoV covers $26\,\mathrm{arcmin}$. The coded-mask imaging technique was adopted as the most suitable approach, given the hard X-ray energy range, large FoV, and localization requirements. This method has been successfully employed in previous space missions, including SIGMA~(\cite{1991AdSpR..11h.289P}), INTEGRAL~(\cite{2003A&A...411L.223G}), and Swift~(\cite{swiftBAT}).

\begin{figure}[ht]
  \centering
   \includegraphics[trim=130 0 0 0, clip, scale=.85]{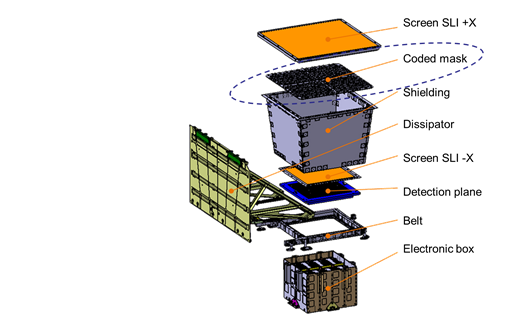}
	  \caption{\label{FigECLAIRs}{\small Parts of the ECLAIRs instrument.} }
\end{figure}

A schematic view of the instrument is shown in Fig.~\ref{FigECLAIRs}. The payload consists of four principal components: (i) a coded mask, (ii) a cooled detection plane comprising 6400 CdTe pixels of $4 \times 4 \times 1\,\mathrm{mm}^3$ active volume and its associated front-end electronics, (iii) a passive lead shielding to ensure that only photons transmitted through the mask are detected, and (iv) dedicated on-board trigger electronics and software (not represented in the figure). Additional technical details on ECLAIRs are provided in \cite{godet25}. The ECLAIRs onboard GRB trigger algorithms based on imaging by coded mask deconvolution are presented in \cite{schanne25}.

The baseline mask parameters were defined in accordance with the ECLAIRs constraints inheriting the satellite and launch vehicle constraints. A square mask with an edge size $L_M = 540\, \mathrm{mm}$ was positioned at a distance of $H = 458.5\,\mathrm{mm}$ above the square detection plane (with an edge size $L_D = 360\, \mathrm{mm}$) to provide the required ECLAIRs FoV ($\sim 2\, \mathrm{sr}$). The allocated mass budget for the mask was limited to approximately $8\,\mathrm{kg}$. Owing to the high complexity and long development timeline, several trade-offs and design iterations were required. 

To enhance sensitivity to highly redshifted GRBs (high-$z$ GRBs), ECLAIRs requires a low-energy threshold of 4~keV while maintaining a 150~keV upper-energy threshold.
The broad band range coverage 4-150~keV has never been achieved in a space-based coded-mask instrument with a wide field of view.
This low-energy constraint introduces significant challenges for both the detection plane and the coded mask. 
Conventional coded masks for hard X-ray/gamma-ray instruments are constructed by mounting opaque elements (typically high-$Z$ materials such as lead) onto a mechanical support structure, such as the composite honeycomb panel used in BAT/Swift~(\cite{swiftBAT}).
The supporting structure is critical during launch, when the mask is subjected to intense vibrations that pose a risk of structural failure. 
However, such designs result in mask “holes” that are not truly open but filled by the support material. This approach is incompatible with a 4~keV threshold, as the supporting material would absorb nearly all X-rays below $\sim 15$~keV — a limitation that sets Swift/BAT low-energy cutoff.

To overcome this limitation, a new coded-mask concept was developed for ECLAIRs. It adopts a self-supporting architecture in which the apertures are fully open and the opaque elements are interconnected, achieving strong mechanical resistance despite the inherent fragility of an open-mask structure.

\section{Pattern generation and selection process}
\label{sect:Pattern}

\subsection{Pattern parameters}

\begin{figure}[htbp]
  \centering
   \includegraphics[trim=300 0 0 0, clip, scale=.064]{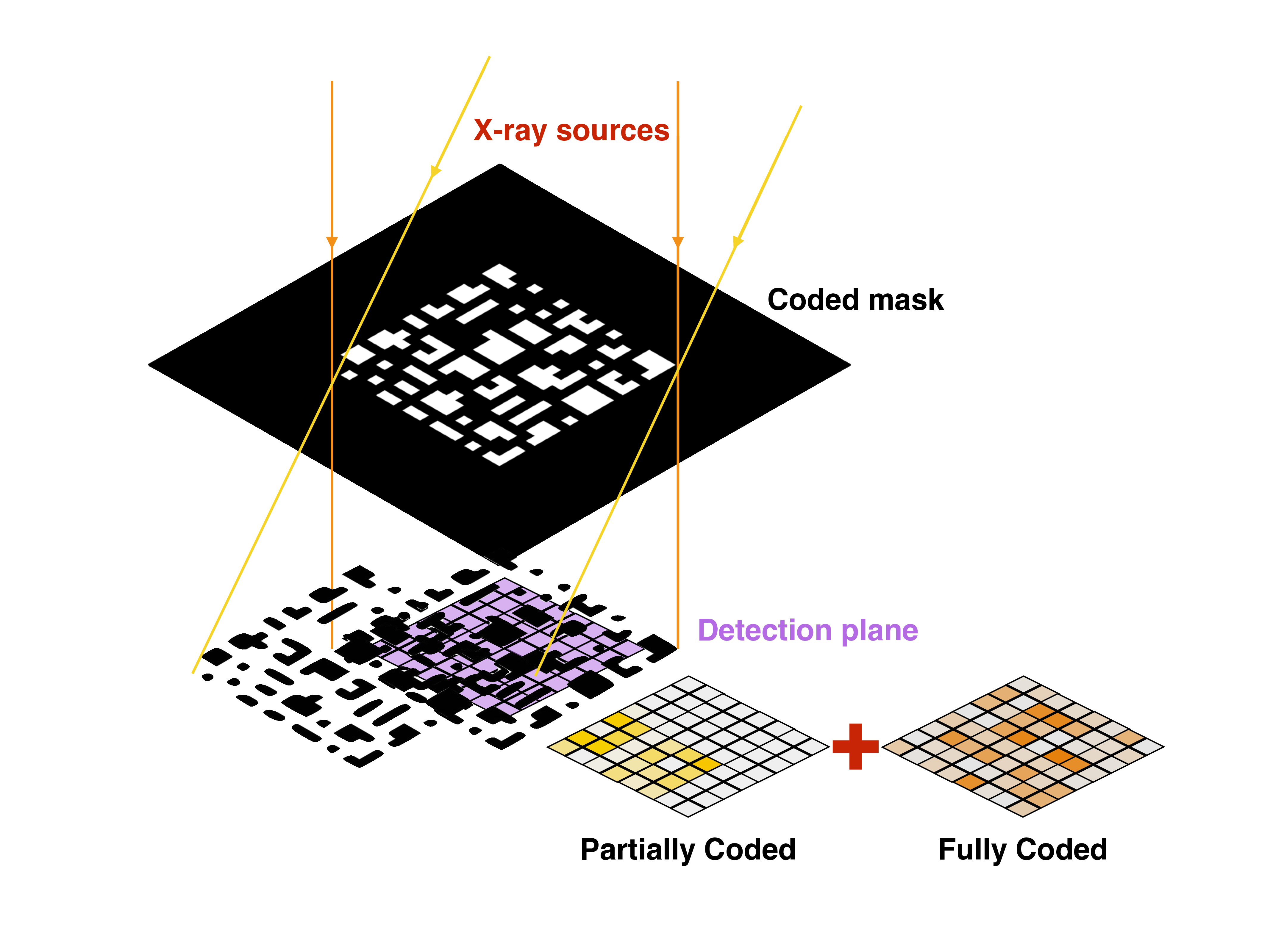}
	  \caption{\label{FigCodedMaskPrinciple}{\small Coded-mask imaging: two X-ray sources project shadowgrams on the detector, shown in orange (fully coded FoV) and yellow (partially coded FoV).} }
\end{figure}

The principle of the coded-mask imaging technique is illustrated in Fig.~\ref{FigCodedMaskPrinciple}.
Each source within the ECLAIRs FoV projects a distinct shadowgram onto the detector plane, where the measured signal corresponds to the superposition of their contributions.
Knowing the mask pattern and the detector pixel counts, the sky image can be reconstructed through deconvolution (\cite{goldwurm25}).

The fully coded FoV corresponds to source directions for which the entire detector is fully coded by the mask, meaning that every detector pixel receives flux modulated by the mask transmission pattern. For sources outside this region, part of the detector is no longer coded by the mask for that direction, so that only a fraction of the detector area contributes to the coded modulation, defining the partially coded FoV.

The fully coded FoV angular aperture is given by $\theta_{FC} = 2 \arctan(\frac{L_M - L_D}{2\,H}) \sim 23\,\mathrm{deg}$ and the partially coded FoV angular aperture (up to zero coding) is given by $\theta_{FC} = 2 \arctan(\frac{L_M + L_D}{2\,H}) \sim 89\,\mathrm{deg}$ \citep{golgro22}. Owing to the large partially coded FoV of ECLAIRs, a random mask pattern was adopted, excluding MURA-type designs to avoid so called "ghost" sources.

The next parameter to be defined is the mask element size, $m$. The detector pixel size is $d = 4.5\,\mathrm{mm}$, including a $0.5\,\mathrm{mm}$ dead zone (\cite{godetDPIX25}) between adjacent detector pixels. The on-axis localization accuracy, expressed by the Point Spread Localization Error (PSLE), can be estimated at the 90~\% C.L. as (see  (\cite{golgro22})):
$$PSLE(SNR) = \arctan \left( \frac{\sqrt{\ln 10}}{SNR}\times \frac{d}{H}\times \sqrt{\frac{m}{d} - \frac{1}{3}} \right)$$
which yields a localization better than 10 arcmin for on-axis sources with $SNR=8$ when $m \sim 120\,\mathrm{mm}$. This value was therefore adopted as a typical mask element size to meet the required localization performance.

In the end, although the mechanical design proved robust—simulations showed that it met the required constraints—a central solid cross was added to the coded mask to provide additional structural margin for this critical element.
Thus the final design is composed of four quadrants of $23\times23$ elements of size $m = 11.394\,\mathrm{mm}$ surrounding a central cross (with size $L_C = m + d = 15.894\,\mathrm{mm}$). We have the equality $L_M = 2 \times 23 \times m + L_C = 47\times m + d$. 
It should be noted that, due to the presence of dead zones between detector pixels, the mask-to-detector pixel size ratio ($m/d$) should be chosen to avoid values close to integers (\cite{golgro22}). For ECLAIRs, the ratios are $m/d = 2.532$ and $L_C/d = 3.532$.

The final parameter to consider is the mask open fraction, defined as the ratio of open area to total mask area. For random masks, the optimum fraction is 50\%. 
However, for a self-supporting mask, a lower open fraction is mechanically easier to achieve, as opaque elements cannot be completely enclosed by transparent elements.
Simulations showed that an open fraction of 40\% affects the sensitivity and localization performance only at the level of a few percent; this value was therefore retained.

\subsection{Self-supporting pattern generation}

\begin{figure}[ht]
  \centering
   \includegraphics[width=0.48\textwidth]{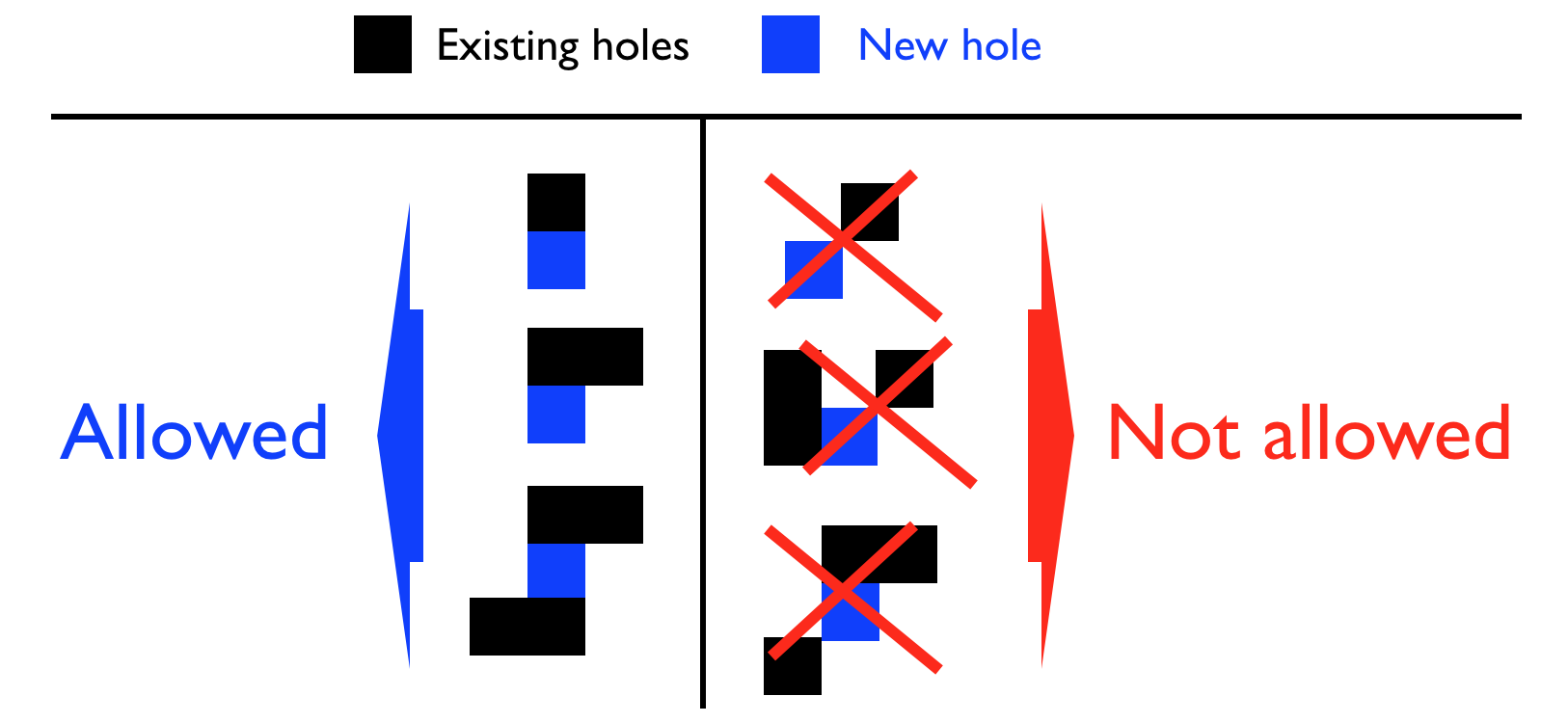}
	  \caption{\label{FigHoleAddition}{\small Rules for accepting or rejecting a new hole, based on the configuration of previously accepted holes, to prevent diagonally connected (corner-only) attachments.} }
    \hfill
\end{figure}

An algorithm was developed to generate mask patterns consistent with the self-supporting requirement, i.e. designs that could be carved directly from a single metal sheet. The pattern was constructed iteratively by removing elements one by one while ensuring that no disconnected structures were created. An additional rule was implemented (see Fig.~\ref{FigHoleAddition}) to prevent neighboring elements from being connected only at their corners, which would constitute a mechanically fragile configuration.

\begin{figure}[ht]
  \centering
   \includegraphics[width=0.37\textwidth]{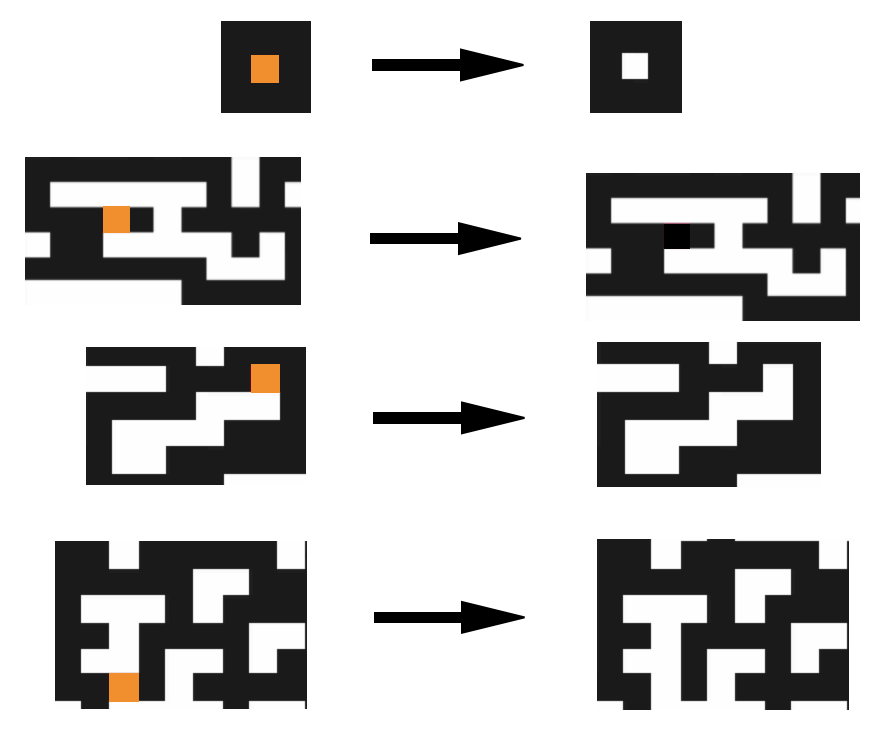}
	  \caption{\label{FigHoleAcceptance}{\small Rules used to accept or reject a new hole, based on the configuration of previously accepted holes (see text for details).} }
\end{figure}

A second objective of the algorithm was to introduce control over the distribution of holes, by grouping them into clusters. Holes are considered part of the same cluster if they are connected, and the relevant physical parameter is the cluster size (CS), defined as the number of elements in a cluster. For a given CS, a hole can be added to an existing cluster only if the resulting cluster does not exceed the specified CS. The acceptance or rejection of a new hole depends on its local configuration, as illustrated in Fig.~\ref{FigHoleAcceptance}, with the following rules:
\begin{enumerate}
\item isolated elements are always accepted,
\item a hole is rejected if its addition would create a disconnected element,
\item a hole is accepted if the resulting cluster remains within the CS limit,
\item if the hole causes the merging of two clusters, the new cluster size must also satisfy the CS constraint.
\end{enumerate}

The pattern generation algorithm introduces a bias near the mask edges, where elements are more likely to find neighbors along the boundary (which is always solid). This results in a non-uniform distribution, with an excess of holes at the edges and a deficit at the center. To mitigate this effect, very large mask patterns ($500 \times 500$ elements) were generated for different CS values and used as a library from which smaller patterns of the required size were extracted, leaving a margin to exclude the edges.

\begin{figure}[!ht]
  \centering
\includegraphics[scale=0.005]{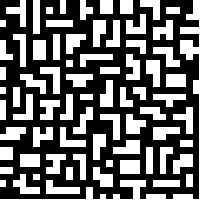}
\includegraphics[scale=0.005]{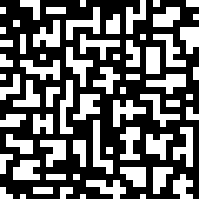}
\includegraphics[scale=0.005]{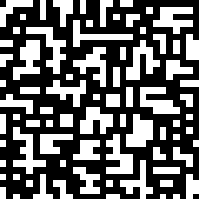}
\includegraphics[scale=0.005]{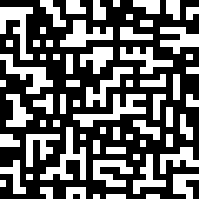}
\includegraphics[scale=0.005]{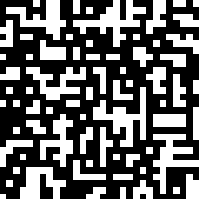}
\includegraphics[scale=0.005]{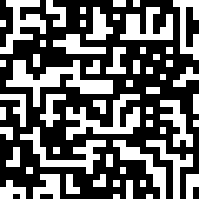}

\caption{Patterns with 40\% open fraction, obtained (from left to right and top to bottom) for $CS = 6, 10, 14, 20, 30,$ and $40$. Each pattern is displayed on a $30 \times 30$ grid.}
\label{figDifferentPatterns}
\end{figure}

Figure~\ref{figDifferentPatterns} illustrates examples of $30 \times 30$ patterns obtained for different cluster sizes. 
It should be noted that an open fraction of 40\% cannot be achieved for $CS \leq 5$.

\subsection{Pattern selection process}

\begin{figure}[ht]
  \centering
   \includegraphics[width=0.36\textwidth]{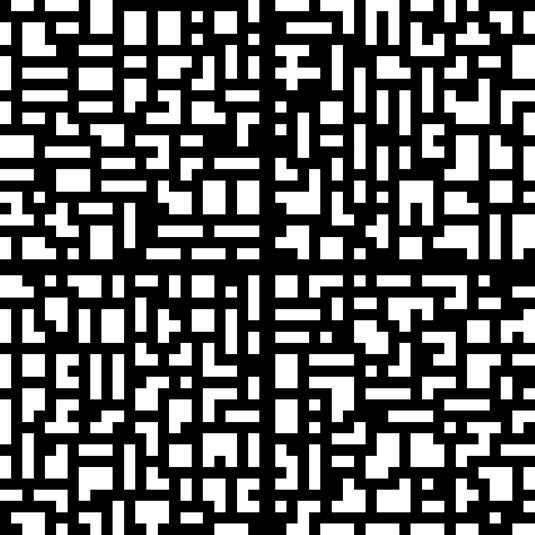}
	  \caption{\label{MaskPattern}{\small Final mask pattern (named C3A2S).} }
\end{figure}

\begin{figure}[ht]
   \includegraphics[width=0.48\textwidth]{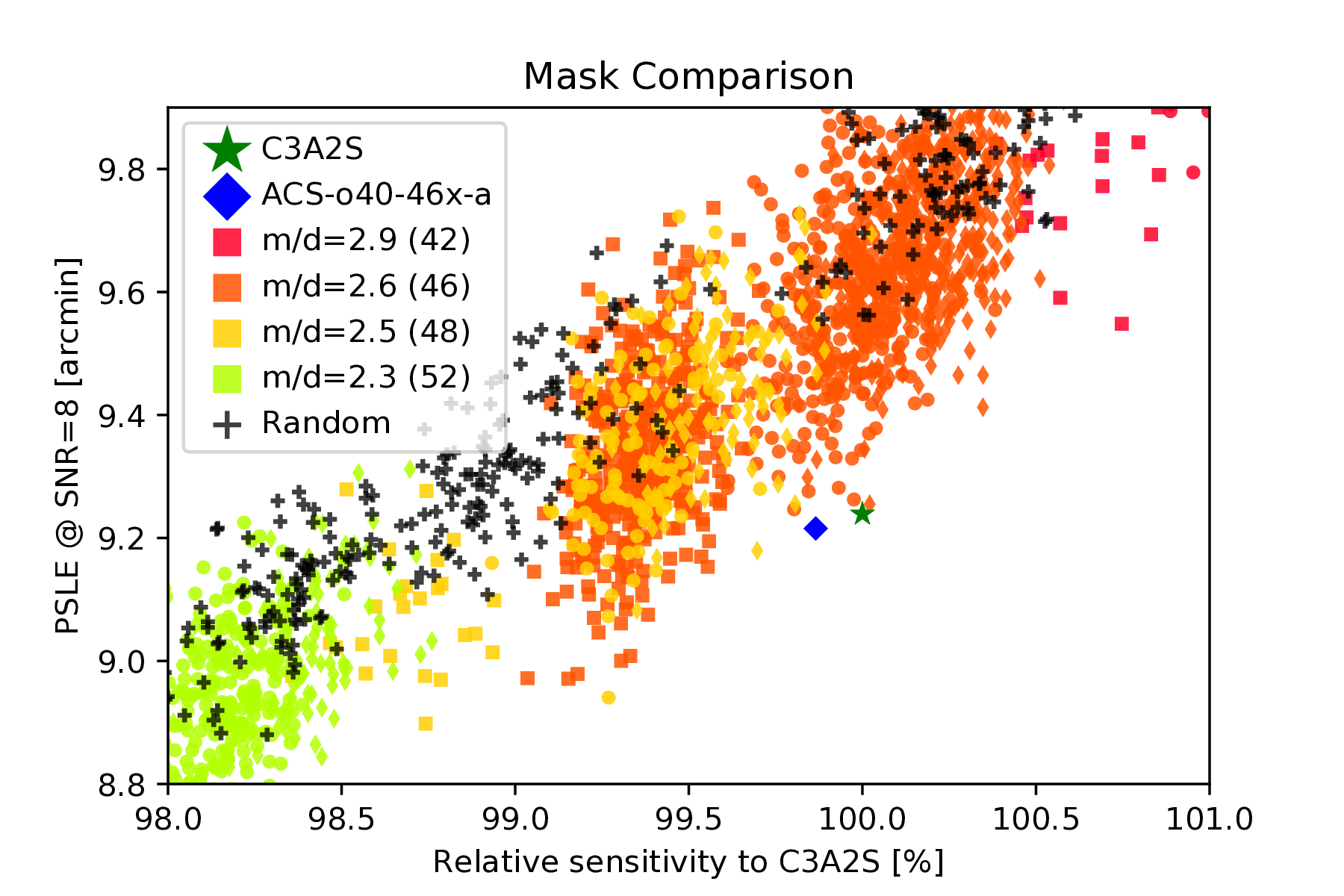}
	  \caption{\label{SNR_PSLE}{\small Localization performance versus sensitivity, using C3A2S as the reference (green star). ACS-o40-46x-a (blue diamond) was the pattern selected prior to the addition of the cross. Colored points correspond to patterns with different $m/d$ values, while purely random masks are shown as black points.} }
\end{figure}

Over 600,000 patterns of size $23 \times 23$ were generated and evaluated in terms of GRB sensitivity and localization performance using a dedicated simulation. From these, the four best candidates were selected and combined to construct the mask pattern.
For mechanical reasons—namely to improve structural resistance (with an overall 20\% gain based on finite element simulations with quasi static loading (QSL) and modal analysis) and to provide sufficient spacing for pin placement—six holes were manually shifted.
This led to the final mask pattern, referred to as C3A2S, shown in Figure~\ref{MaskPattern}. The localization performance of the generated patterns as a function of sensitivity is presented in Figure~\ref{SNR_PSLE}, taking C3A2S as the reference. As illustrated, C3A2S indeed provides the best localization accuracy for a given sensitivity.

\subsection{System Point Spread Function}
\begin{figure}[ht]
  \centering
   \includegraphics[width=0.48\textwidth]{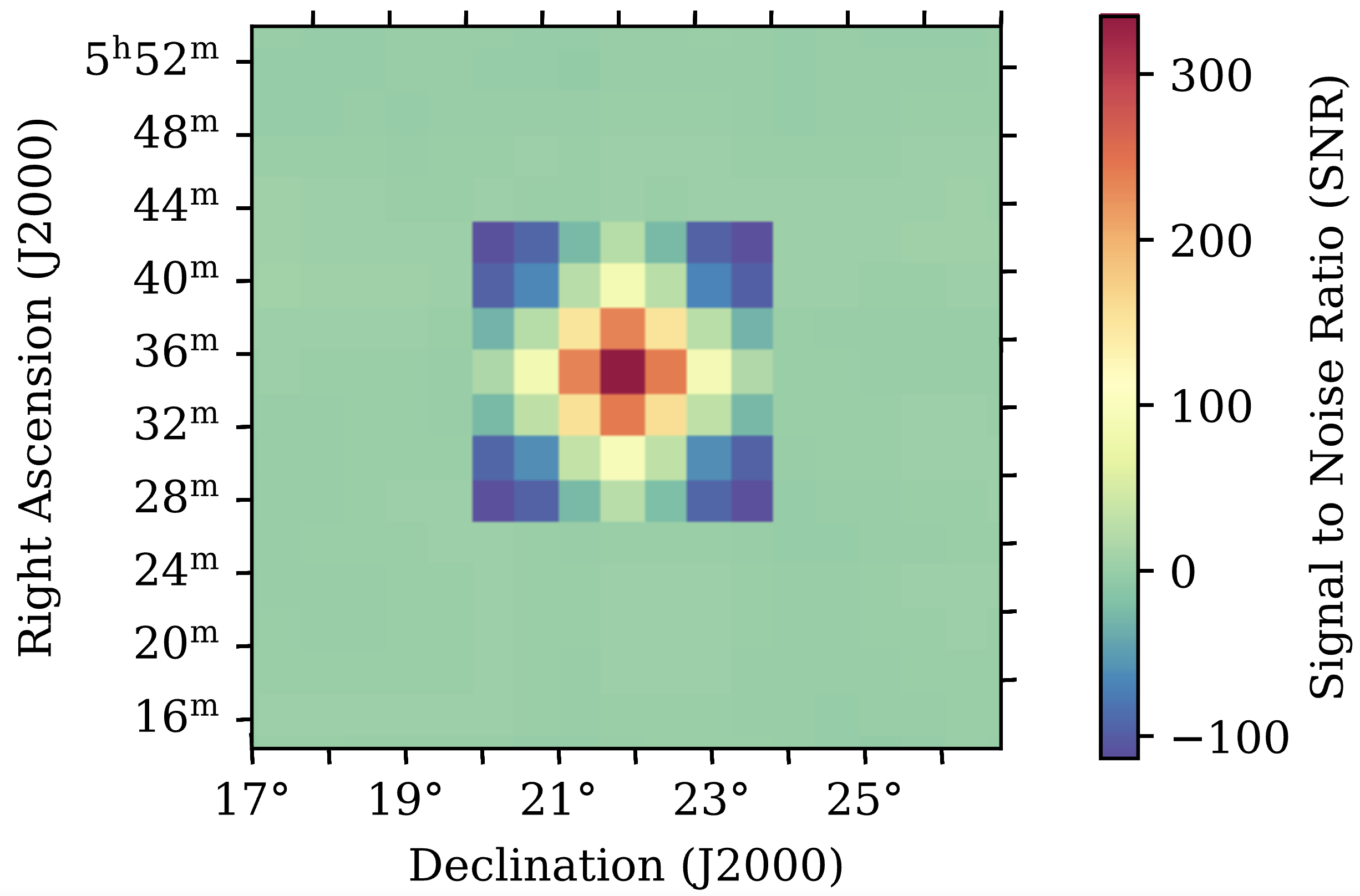}
	  \caption{\label{PSF}{\small Sky image in signal-to-noise ratio (SNR) centered on the Crab (for the low energy band 4-10~keV). The central peak is surrounded by coding noise with alternating positive and negative lobes, originating from algorithm used to construct the mask pattern.} }
\end{figure}

The System Point Spread Function (SPSF) of the coded mask system, \it i.e. \rm  the response of the imaging system to a point source after deconvolution (\cite{golgro22}), resulting from the pseudo-random coded mask pattern we developed, exhibits a characteristic structure. Because the algorithm penalizes diagonal elements relative to lateral ones (see Fig.~\ref{FigHoleAddition}), the central source peak is surrounded by a “coding noise,” consisting of numerous small secondary positive and negative lobes (Figure~\ref{PSF}).

\section{Mechanical design}
\label{sect:Discussion}

The main mechanical requirements for the coded mask are as follows:
\begin{itemize}
    \item thermal endurance over the operational range of $-60^{\circ}$C to $+55^{\circ}$C.
    \item a first eigenfrequency above 120 Hz,
    \item compliance with QSL, sinusoidal, and random vibration loadings (see table~\ref{tab:vibration}), whose levels increased significantly during development (since the mask, being located at the top of the payload, is subject to the cumulative dynamic load responses of every structures underneath
\end{itemize}

\begin{table}[h]
\centering
\begin{tabular}{lccc} 
\hline
 & \\[\dimexpr-\normalbaselineskip+2pt]
 & X Axis & Y Axis & Z Axis \\
 \hline
 
 & \\[\dimexpr-\normalbaselineskip+2pt]
 
QSL & 77g & 29g & 39g \\
Sinus ($0\rightarrow100\,Hz$) & 24g & 30g & 26g \\
Random& 37grms & 13grms & 21grms \\
\hline
\end{tabular}
\caption{Vibration load levels used in the qualification tests of the coded mask. X Axis is perpendicular to the mask. Y and Z Axis are on the mask plane.}
\label{tab:vibration}
\end{table}

\begin{figure}[htbp]
  \centering
  
   \includegraphics[width=0.48\textwidth]{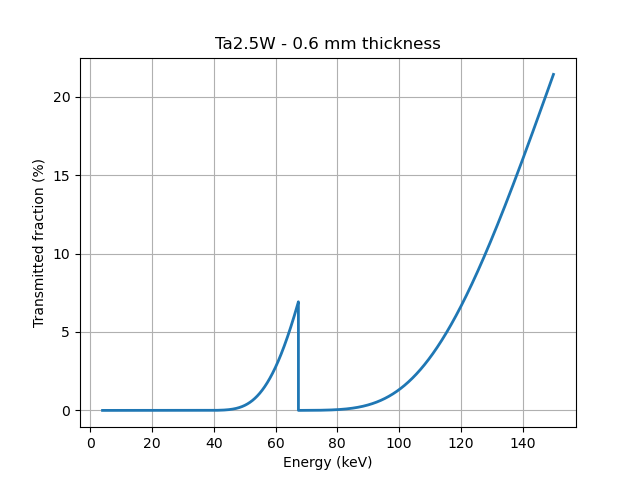}
	  \caption{\label{TransmittedFraction}{\small 
      Photon transmission as a function of energy over the 4–150 keV ECLAIRs band for a 0.6 mm thick Ta2.5W material.} }
\end{figure}

\begin{figure*}[hbtp]
  \centering
\includegraphics[width=.8\textwidth]{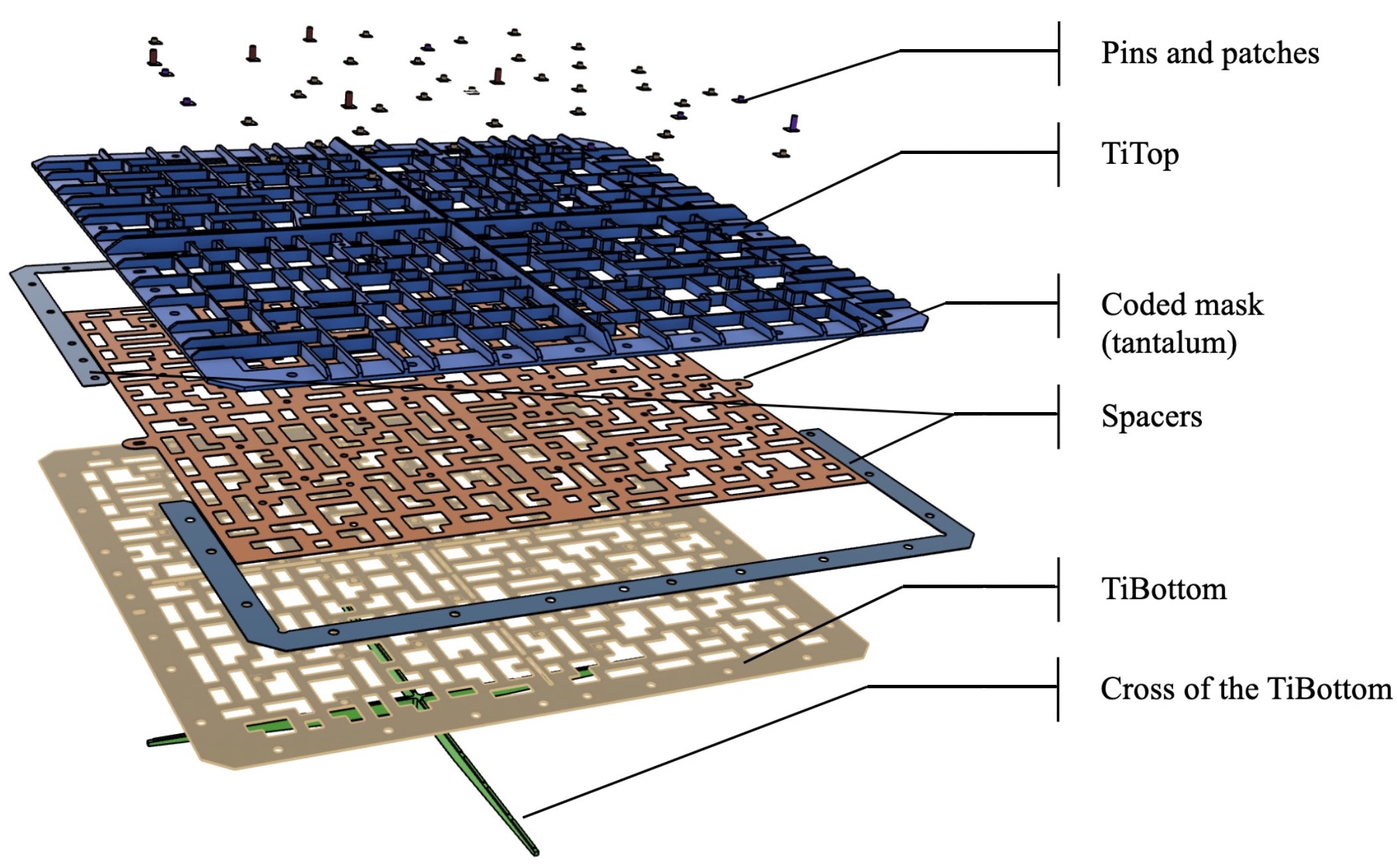} 
\caption{Exploded view of the coded mask assembly.}
\label{figMaskAssembly}
\end{figure*}

In its nominal dimensions, the mask would exhibit a first eigenfrequency of approximately 15 Hz (\cite{2024SPIE13093E..70G}). To increase this value, several approaches were investigated, including stretched tantalum sheets and multilayer titanium/silicone/tantalum/silicone/titanium sandwich structures. These options were ultimately discarded because of the more stringent vibration requirements introduced during development. The adopted solution was to incorporate stiffening elements perpendicular to the mask surface in order to enhance its structural rigidity. The resulting design consists of a $0.6\,\mathrm{mm}$ thick tantalum alloy (Ta2.5W) sheet, providing sufficient opacity to X-rays over the 4–150 keV energy range
(figure~\ref{TransmittedFraction}) and is thin enough to reduce significantly the vignetting effect, sandwiched between two titanium (TA6V alloy, also known as Ti-6Al-4V, offering excellent yield strength and a coefficient of thermal expansion close to that of Ta2.5W) stiffeners, referred to as \textit{TiTop} and \textit{TiBottom}. The complete assembly is denoted as \textit{TiTaTi} (figure~\ref{figMaskAssembly}).

\begin{figure}[ht]
  \centering
\includegraphics[width=0.48\textwidth]{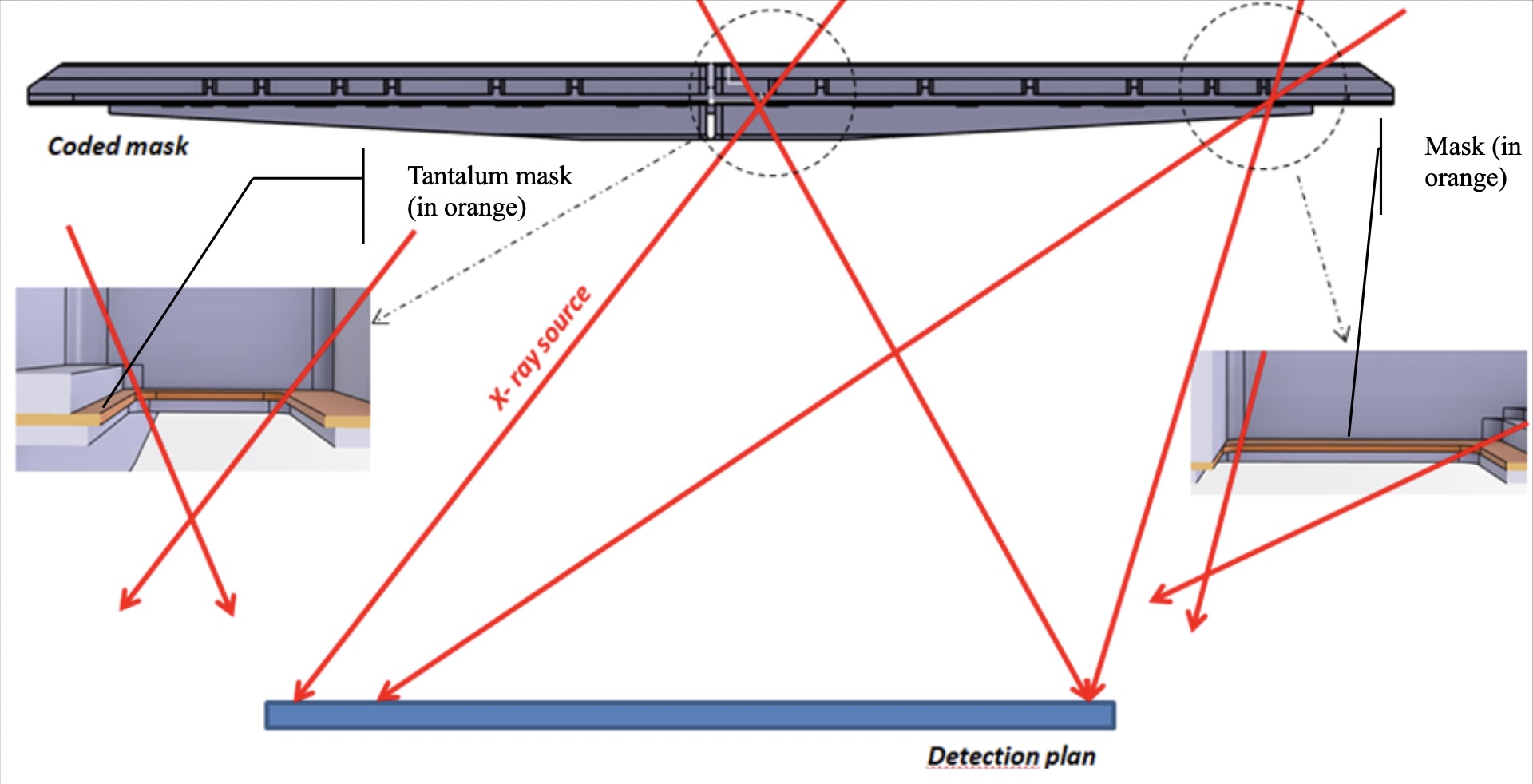}
\caption{Design of the \textit{TiTop} and \textit{TiBottom}. The stiffeners are positioned so as not to interfere with the imaging process; for the coded-mask principle, only the tantalum sheet pattern needs to be considered.}
\label{figTiTaTi}
\end{figure}

The tantalum sheet is fixed at a single point and otherwise allowed to expand freely under temperature variations, with an oblong constraint guiding the thermally induced expansion, thereby avoiding mechanical stress and minimizing deviations from planarity.
The \textit{TiTop} and \textit{TiBottom} plates each feature a central cross of $15\,\mathrm{mm}$. To further enhance mechanical resistance, the \textit{TiTop} is reinforced with $10\,\mathrm{mm}$ stiffeners all around the pattern. The \textit{TiTop} component is machined from a single 2\,cm-thick TA6V block through successive material-removal steps. The \textit{TiBottom} cross is likewise produced from a monolithic TA6V block and subsequently welded onto the \textit{TiBottom} plate under a nitrogen atmosphere to minimize distortion. The overall design ensures that photons within the ECLAIRs FoV can only intersect the \textit{TiTop} or \textit{TiBottom} if they also pass through the tantalum layer (see Figure~\ref{figTiTaTi}). For sky image reconstruction, the mask pattern of the tantalum sheet only is to be considered.

\begin{figure}[!ht]
  \centering
\includegraphics[width=0.48\textwidth]{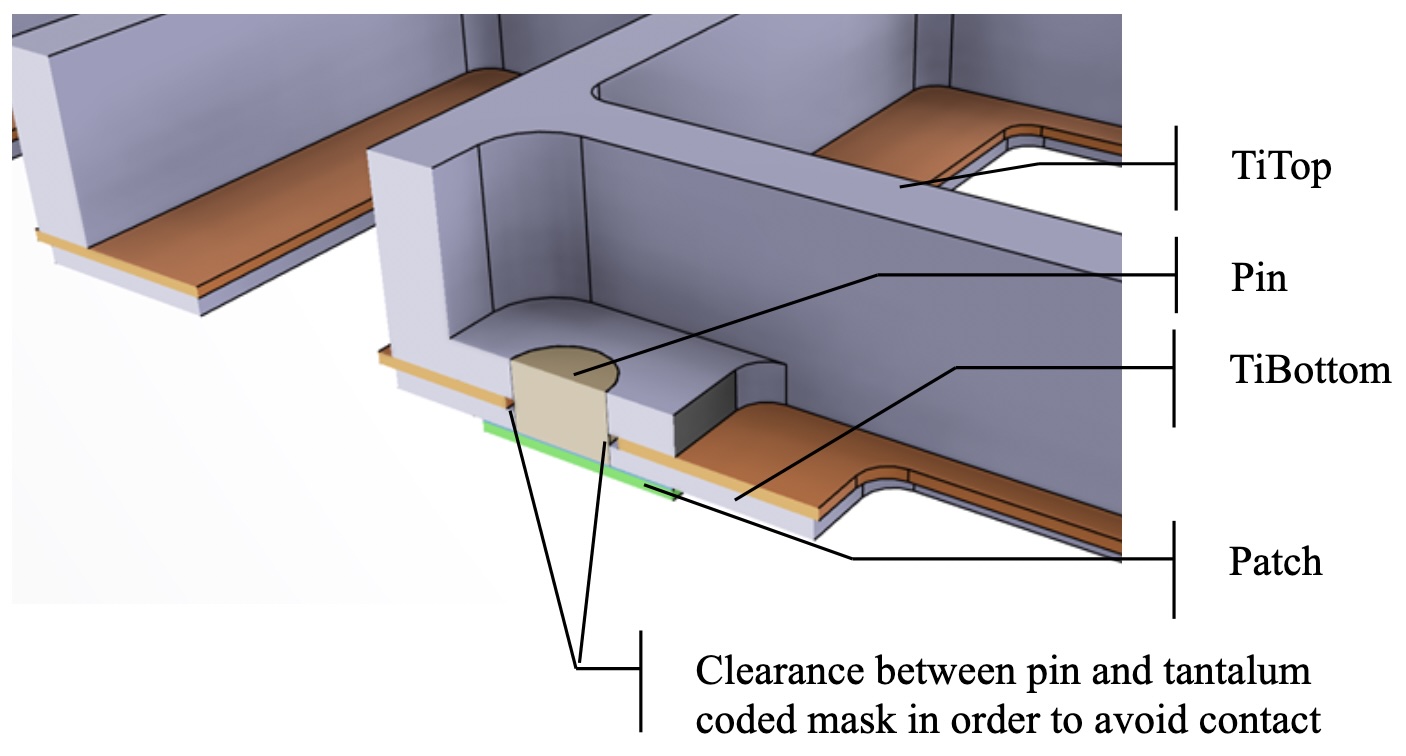}
\caption{Pins used to assemble the \textit{TiTop} and \textit{TiBottom}; tantalum patches are added to block photons from passing through the pin holes.}
\label{figPinsPatches}
\end{figure}

The \textit{TiTop} and \textit{TiBottom} are assembled using 45~pins (TA6V) and laser YAG welding. The tantalum sheet is perforated to allow the pins to pass through without direct contact (to allow thermal expansion). To prevent photons from passing through these pin holes and reaching the detector plane, the openings in the tantalum are sealed with tantalum patches, which are glued onto the \textit{TiBottom} (figure~\ref{figPinsPatches}).

Thanks to this complex design, the mask successfully passed all mechanical and thermal qualification tests. A first vibration mode of $145\,\mathrm{Hz}$ was measured, fully compliant with the primary requirement.

\begin{figure}[ht]
  \centering
   \includegraphics[width=0.48\textwidth]{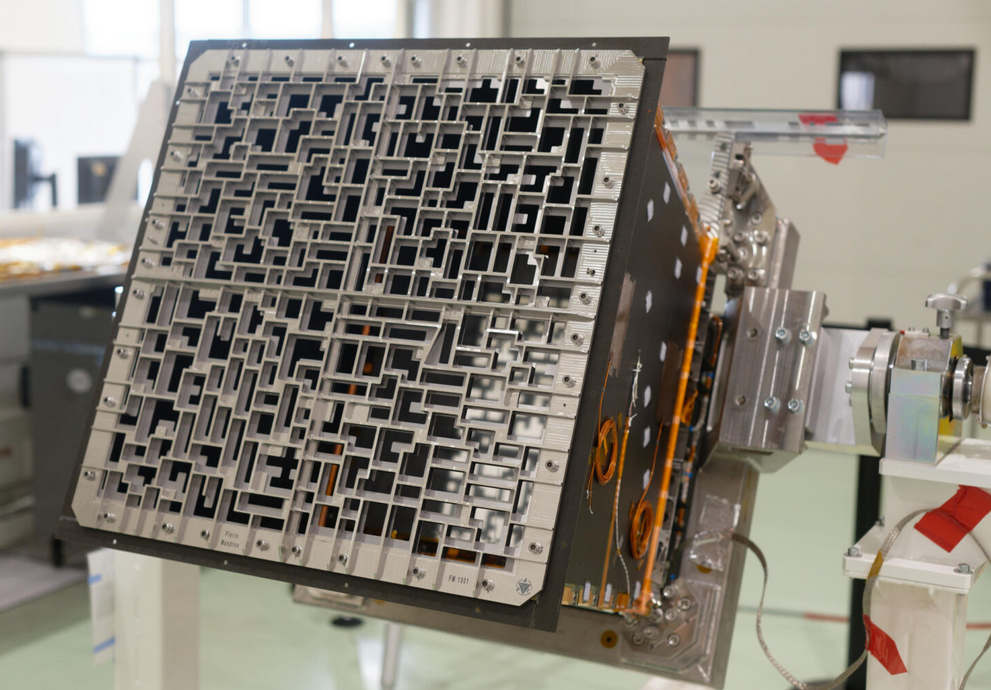}
	  \caption{\label{CodedMaskFinal}{\small The coded mask installed on top of the ECLAIRs camera before the mask is covered with the insulation protection. Credit : CNES.} }
\end{figure}

The fabrication of a single $7.4\,\mathrm{kg}$ mask required more than $80\,\mathrm{kg}$ of raw materials. Figure \ref{CodedMaskFinal}
 shows the coded mask flight model installed on top of the ECLAIRs instrument.

\section{Meteorites and detector protection}

An inherent consequence of the self-supporting mask design is the direct exposure of the detector plane to the space environment. Since the detectors must operate in darkness, an additional protective layer is required, as thin as possible so as not to compromise the $4\,\mathrm{keV}$ low-energy threshold (a single-layer insulation was initially considered). To mitigate the risk of micrometeorite impacts creating holes in this layer, the final design incorporates two layers: one positioned directly above the detector plane, and a second placed on top of the mask (figure \ref{FigECLAIRs}) such that a  puncture of both layers by a micrometeorite only affect a single sky direction and not the whole FoV.

\section{First light in space environment}

\begin{figure}[!ht]
  \centering
   \includegraphics[width=0.4\textwidth]{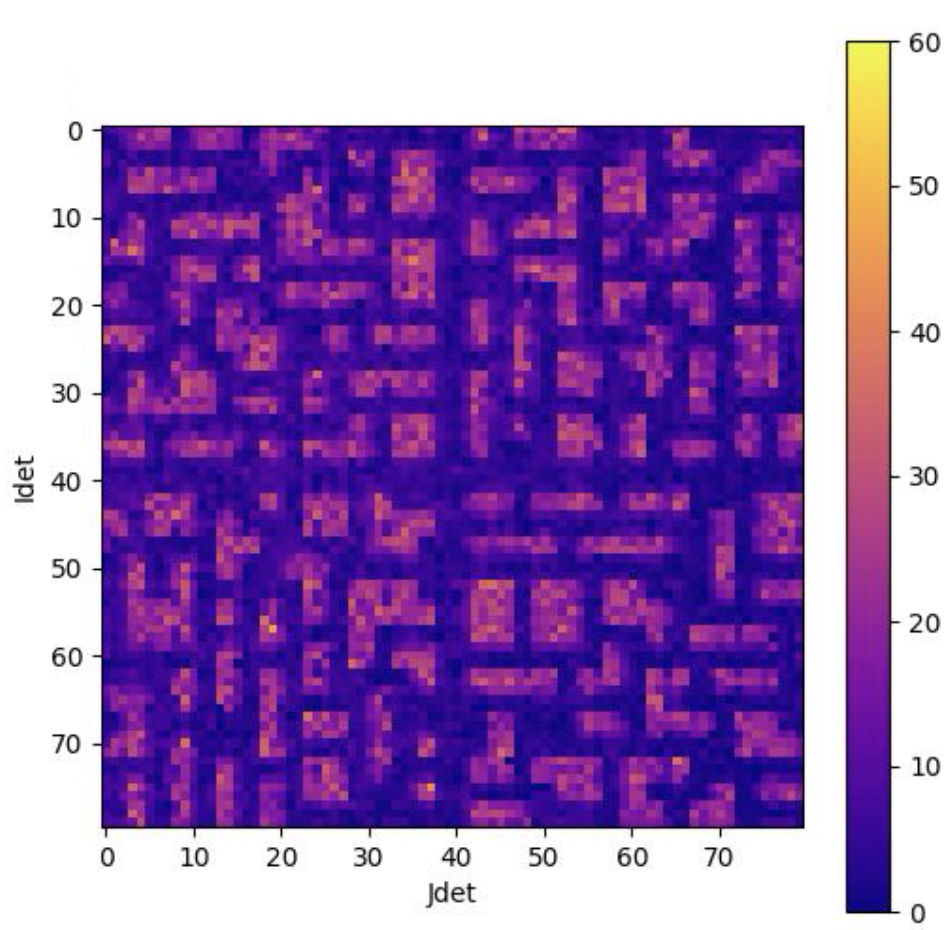}
	  \caption{\label{FirstShadowgram}{\small First light recorded by the detector plane (Idet and Jdet are pixels coordinates) while ECLAIRs points at Sco X-1. The pattern of the coded mask is visible in this first shadowgram.} }
\end{figure}

SVOM was launched on 22 June 2024 from Xichang (China) aboard a Long March 2C rocket. Following completion of the early orbital phase, the ECLAIRs instrument was switched-on on the 5th of July and pointed toward the bright soft X-ray source Sco~X-1. The detector plane successfully recorded its first light, producing a clear shadowgram that confirmed the mask had withstood launch and was performing as expected (see figure \ref{FirstShadowgram},  the central mask cross shows good alignment of the source with the ECLAIRs optical axis). See (\cite{goldwurm25}) for a deconvolved image of this very first Sco X-1 observation.

\section{Conclusions}
\label{sect:Conclusion}

To achieve the low-energy threshold of $4\,\mathrm{keV}$ while maintaining a 150~keV upper-energy threshold, it was necessary to develop a new type of coded mask based on a self-supporting design. The construction of such a mask posed both scientific and mechanical challenges. Several innovations were introduced, including dedicated generative algorithms for pattern design and a sandwich structure with stiffeners that preserved imaging performance. 
 This achievement required twelve years of development at the APC laboratory, in close collaboration with CNES and CEA.

\begin{acknowledgements}
The Space-based multi-band astronomical Variable Objects Monitor (SVOM) is a joint Chinese-French mission led by the Chinese National Space Administration (CNSA), the French Space Agency (CNES), and the Chinese Academy of Sciences (CAS). We gratefully acknowledge the unwavering support of NSSC, IAMCAS, XIOPM, NAOC, IHEP, CNES, CEA, 
CNRS and Université Paris Cité
\end{acknowledgements}

\label{lastpage}

\end{document}